# Future High Energy Circular e+e- Collider using Energy-Recovery Linacs

Vladimir N Litvinenko[1,2], Thomas Roser[2] and Maria Chamizo Llatas[3]

[1] Department of Physics and Astronomy, Stony Brook University, Stony Brook, NY, USA

[2] Collider-Accelerator Department, Brookhaven National Laboratory, Upton, NY, USA

[2] Nuclear and Particle Physics Directorate, Brookhaven National Laboratory, Upton, NY, USA

*In this paper we present alternative approach for Future Circular electron-positron Collider. Current 100 km circumference design with the top CM energy of 365 GeV (182.5 GeV beam energy) is based on two storage rings to circulate colliding beams [1-2]. One of the ring-ring design shortcomings is enormous power consumption needed to compensate for 100 MW of the beam energy losses for synchrotron radiation. We propose to use energy recovery linac located in the same tunnel to mitigate this drawback. We show in this paper that our approach would allow a significant – up to an order of magnitude – reduction of the beam energy losses while maintaining high luminosity in this collider at high energies. Furthermore, our approach would allow to extend CM energy to 500 GeV (or above), which is sufficient for double-Higgs production.*

*Introduction.* The current ring-ring design of the Future Circular electron-positron Collider (FCC ee) (see[1-3,6] and references therein) aims to achieve the top CM energy of 365 GeV with a luminosity of 1.5-3x10$^{34}$ cm$^{-2}$s$^{-1}$ using 100 MW of RF power compensating for the synchrotron radiation of the electron and position beams, which likely would result in a wall-plug AC power of 200MW. At lower energies, with the same level of RF power, the FCC ee luminosity would grow approximately as E$^{-3.6}$. While the ring-ring FCC ee promises a very high luminosity of 4.5x10$^{36}$ cm$^{-2}$s$^{-1}$ at CM energy of 91.3 GeV, it drops more than two orders of magnitude to 3x10$^{34}$ cm$^{-2}$s$^{-1}$ at CM energy of 365 GeV.

In this paper we propose another approach to the FCC ee based on colliding electron and positron beams accelerated and decelerated in an Energy-Recovery Linac (ERL) located in the same FCC tunnel with 100 km circumference. Our approach is a natural extension of that developed for an ERL-based electron-ion collider at Brookhaven National Laboratory (eRHIC) where a 20 GeV electron beam collides with a 275 GeV proton beam [4-5]. Use of this approach indicates that an ERL-based FCC ee can reach significantly higher energy as well as higher luminosity, when compared with the existing ring-ring design, while significantly reducing the required RF power.

*ERL-based FCC ee scheme.* The relation between the required RF power to compensate for the beam energy losses from synchrotron radiation in an accelerator is

$$P_{SR} = V_{SRe-} I_{e-} + V_{SRe+} I_{e+} \qquad (1)$$

where $eV_{SRe-}, eV_{SRe+}$ are synchrotron radiation (SR) beam energy losses by electrons and positrons (*e* is the charge of positron) and $I_{e-}, I_{e+}$ are the electron and positron beam currents. The collider luminosity is then given by

$$L = f_c \frac{N_{e-}N_{e+}}{4\pi\sqrt{\beta_x^*\varepsilon_x}\sqrt{\beta_y^*\varepsilon_y}} h = \frac{I_{e-}I_{e+}}{4\pi\sqrt{\beta_x^*\varepsilon_x}\sqrt{\beta_y^*\varepsilon_y}\cdot f_c \cdot e^2} h \qquad (2)$$

where $\beta_{x,y}^*$ are the β-functions at the collision point, $\varepsilon_{x,y}$ are the horizontal and vertical geometric RMS emittances, $f_c$ is the bunch collision energy and $h\sim 1$ is so-called hour-glass effect. With a fixed RF power and equal SR losses $V_{SRe-} = V_{SRe+} = V_{SR}$, the maximum luminosity is attained with equal beam currents $I_{e-} = I_{e+} = I$

$$L = \frac{h}{16\pi\sqrt{\beta_x^*\varepsilon_x}\sqrt{\beta_y^*\varepsilon_y}\cdot f_c}\left(\frac{P_{SR}}{eV_{SR}}\right)^2. \qquad (3)$$

Hence, the only way of increasing luminosity at a given power consumption is reducing the denominator

$$f_c \cdot \sqrt{\beta_x^*\beta_y^*}\cdot\sqrt{\varepsilon_x\varepsilon_y}\ . \qquad (4)$$

In a ring-ring collider such an optimization is limited by the allowable tune-shift in the beam-beam collisions:

$$\xi_{x,y} = \frac{Nr_e\beta_{x,y}^*}{2\pi\sigma_{x,y}(\sigma_x+\sigma_y)} = \frac{Nr_e\sqrt{\beta_{x,y}^*}}{2\pi\sqrt{\varepsilon_{x,y}}\left(\sqrt{\varepsilon_x\beta_x^*}+\sqrt{\varepsilon_y\beta_y^*}\right)} \le \xi_{max};\ \sigma_{x,y}\equiv\sqrt{\beta_{x,y}^*\varepsilon_{x,y}}, \qquad (5)$$

which limits the reduction of the beam emittances $\varepsilon_{x,y}$ and bunch collision frequency $f_c$ to those chosen for FCC ee ring-ring design. Similarly, values of $\beta_{x,y}^*$ are already optimized for the FCC ee ring-ring design and a further reduction is very unlikely[1].

In contrast, collisions of electron and positron beams accelerated either in linear accelerators (linacs) or energy-recovery linacs (ERL) are no longer limited by the tune-shift condition [4]. Removing these restrictions allows further optimization of the FCC ee luminosity[2].

At FCC energies ERLs have significant advantages, when compared with linear colliders, by both recovering significant portions of the beam energy as well as recycling both electron and positron beams in relatively low energy (~ 2 GeV) storage rings used for cooling the beams. As a result, the ERL-based FCC ee promises to consume significantly less power while providing higher luminosities at energies of interest for FCC ee. While such a design is not capable to compete with the very high ring-ring luminosity at CM energy of 92 GeV (Z pole), if proven, it could deliver higher or comparable luminosities at CM energies above 150 GeV. Furthermore, it could extend the FCC ee energy reach to the double-Higgs production energy of 500 GeV or, if needed, slightly higher. Figure 1 shows a comparison of our luminosity estimations for a ERL-based FCC ee with other options. The most remarkable promise of the ERL-based FCC ee collider is the possibility of delivering high luminosity at high energies, while consuming a small portion (~10%) of the RF power, when compared with the ring-ring design. It turns out that an ERL-based ee collider might also deliver higher luminosity with lower electric power consumption in the lower energy range of linear colliders.

---

[1] Beam parameter for the ring-ring FCC ee design, subject to a variety of constrains, are indeed very well optimized – see [1-3] for details.
[2] In addition, if desired, such collider can collide polarized electron and position beams with high degree of polarization

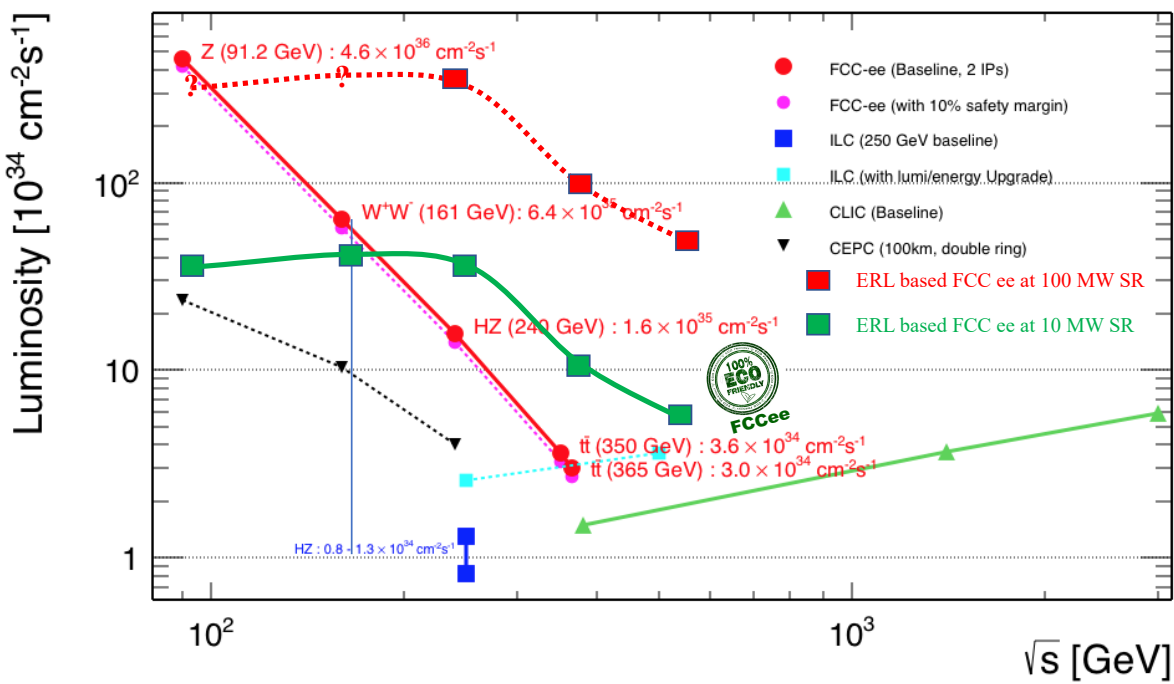

Fig. 1. Luminosities for various options of an FCC ee. The original plot for the FCC ee ring-ring design and other colliders is taken from [6]. The thick green line and green squares show our estimated luminosities for the ERL-based collider consuming 10 MW of RF power, the design we call **Green FCC ee**. The red dash-line shows a simple linear scaling of the luminosities to 100 MW RF – this mode is not what we are proposing for the FCC ee ERL-based design.

A possible realization for an ERL-based FCC ee collider is shown in Fig. 2. Low emittance flat electron and positron beams from 2 GeV cooling rings are injected into and accelerated to the top energy in a multi-turn ERL (see Table 1 for details) comprised of two superconducting RF (SRF) linacs located in the FCC tunnel. While beams with intermediate energies bypass the interaction regions, beams at the top energy do collide in one of the interaction regions (IRs). A relatively low bunch repetition rate (see Table 1) allows one to time individual bunches so that they collide in one of the IRs. In this scenario the luminosity can be divided (shared) between the IRs in any desirable ratio[3]. The used beams, with significantly increased energy spreads and emittances, are then decelerated in the ERL to 2 GeV, reinjected into the storage rings and cooled there to the required low emittances before repeating the trip in the ERL. Beam losses, which are expected to be very low, are replenished by top-off injection from two 2 GeV linacs equipped with electron and position sources[4].

---

[3] The other scenario, when beams can collide in each IR with increases overall luminosity is also possible, but it required detailed studies elsewhere

[4] If desired, these could be polarized electrons and positrons

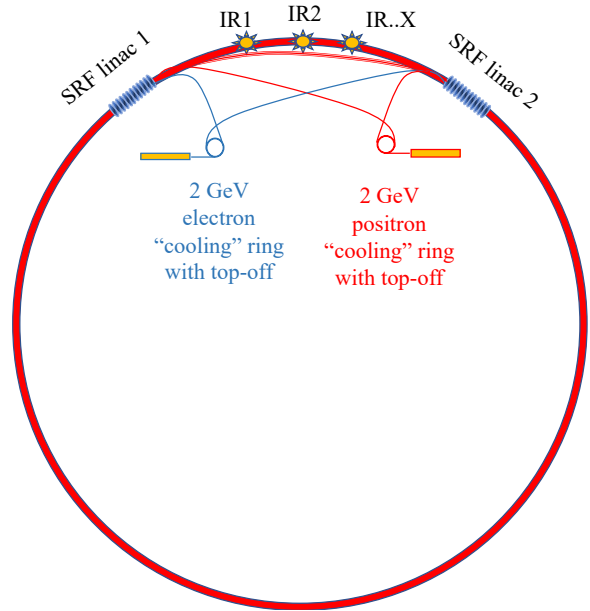

Fig. 2. A possible options of an ERL-based FCC ee collider with linacs separated by 1/6$^{th}$ of the FCC circumference.

| FCC with ERLs | Z | W | H(HZ) | ttbar | HH |
|---|---|---|---|---|---|
| *Circumference, km* | *100* | *100* | *100* | *100* | *100* |
| **Beam energy, GeV** | **45.6** | **80** | **120** | **182.5** | **250** |
| Horizontal norm ε, μm rad | 4 | 4 | 6 | 8 | 8 |
| Vertical norm ε, nm rad | 8 | 8 | 8 | 8 | 8 |
| *βh, m (same as in FCCee design)* | 0.15 | 0.2 | 1 | 1 | 1 |
| *βv, mm same as in FCCee design)* | 0.80 | 1.00 | 1.00 | 2.00 | 2.00 |
| **Bunch length, mm** | **0.8** | **1** | **1** | **2** | **2** |
| Charge per bunch, nC | 12.5 | 12.5 | 25 | 22.5 | 19 |
| Ne per bunch | 7.8E+10 | 7.8E+10 | 1.6E+11 | 1.4E+11 | 1.2E+11 |
| **Bunch frequency, kHz** | **99** | **90** | **33** | **15** | **6** |
| **Beam current, mA** | **1.24** | **1.12** | **0.82** | **0.34** | **0.11** |
| **Luminosity, $10^{34}$ cm$^{-2}$sec$^{-1}$** | **22.5** | **28.9** | **25.9** | **10.5** | **4.5** |

Table 1. Main parameters of possible ERL-based FCC ee

Naturally, the ERL will not recover all of the beam energy – at the top FCC ee energies a significant portion of the beam energy will be lost to synchrotron radiation. Furthermore, since the ERL beams are passing around the FCC tunnel on their way-up in energy and on their way down, synchrotron losses for a cycle (from the 2 GeV cooler ring up and returning back) exceed those of a single path in a ring. As can be seen in Table 2, SR losses, which include SR power in the damping rings, increase with the number of ERL passes, while the required linac voltage is reduced. However, it is unlikely that increasing the number of passes beyond 6 would have any advantages.

| FCC with ERLs | Z | W | H(HZ) | ttbar | HH |
|---|---|---|---|---|---|
| Beam energy, GeV | 45.6 | 80 | 120 | 182.5 | 250 |
| **Four path ERL + Damping ring** | | | | | |
| Energy loss per particle, GeV | 4.0 | 4.4 | 6.0 | 14.8 | 42.7 |
| Radiated power, MW/per beam | 5.0 | 5.0 | 5.0 | 5.0 | 4.9 |
| ERL linacs voltage, GV | 10.88 | 19.6 | 29.8 | 46.5 | 67.4 |
| **Six path ERL + Damping ring** | | | | | |
| Energy loss per particle, GeV | 4.1 | 4.6 | 7.1 | 20.4 | 64.5 |
| Radiated power, MW/per beam | 5.0 | 5.2 | 5.9 | 6.9 | 7.4 |
| ERL linacs voltage, GV | 7.25 | 13.1 | 20 | 31.6 | 47.7 |
| **Secondary parameters** | | | | | |
| Disruption, Dx | 0.6 | 0.6 | 0.1 | 0.2 | 0.2 |
| Disruption, Dy | 183 | 177 | 129 | 143 | 121 |
| Energy loss in IP, GeV | 0.05 | 0.16 | 0.28 | 0.30 | 0.55 |
| Tune shift, χ hor | 8.9 | 8.9 | 11.7 | 8.0 | 6.8 |
| Tune shift, χ ver | 14.5 | 14.1 | 10.2 | 11.3 | 9.6 |
| **Cooler rings** | | | | | |
| Cooler ring energy, GeV | 2 | 2 | 2 | 2 | 2 |
| Damping time, msec | 2.0 | 2.0 | 2.0 | 2.0 | 2.0 |
| Beam current, mA | 534 | 486 | 356 | 146 | 49 |

Table 2. FCC ee parameters for 4-path and 6-path ERL.

In an ERL, on the way up in energy, both the electron and positron bunches pass the ERL linacs in the accelerating phase (on-crest or close to it), as shown in Figure 3. At the top energy the phase for electrons and positron changes from accelerating to decelerating by acquiring an additional 180-degrees shift and beams give energy back to the cavities. When losses from synchrotron radiation are relatively small [4-5] the beam energies on the way up and down are nearly symmetric: at corresponding passes the energies of the accelerating and decelerating beams are close to each other and the same magnetic system can be used to transport them around the ring tunnel.

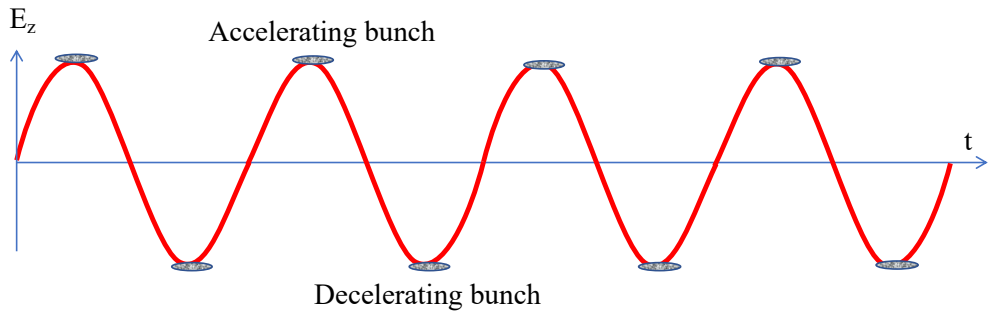

Fig. 3. Illustration of ERL principle (intentionally simplified): accelerating bunches take energy from SRF linac, while decelerating bunches return energy back.

Figure 4 shows examples of the energy evolution in the ERL for 4 and 6 passes with top beam energies of 182.5 GeV and 250 GeV (CM energies of 365 GeV and 500 GeV). At top FCC ee energies synchrotron losses per pass are significant and such symmetry as in Fig. 3 is lost. As can be seen in the beam energy tracking plots in Figure 4, the SR losses make the energy evolution asymmetric with respect to the top energy, e.g. energies in the decelerating pass differ significantly from those on the way up.

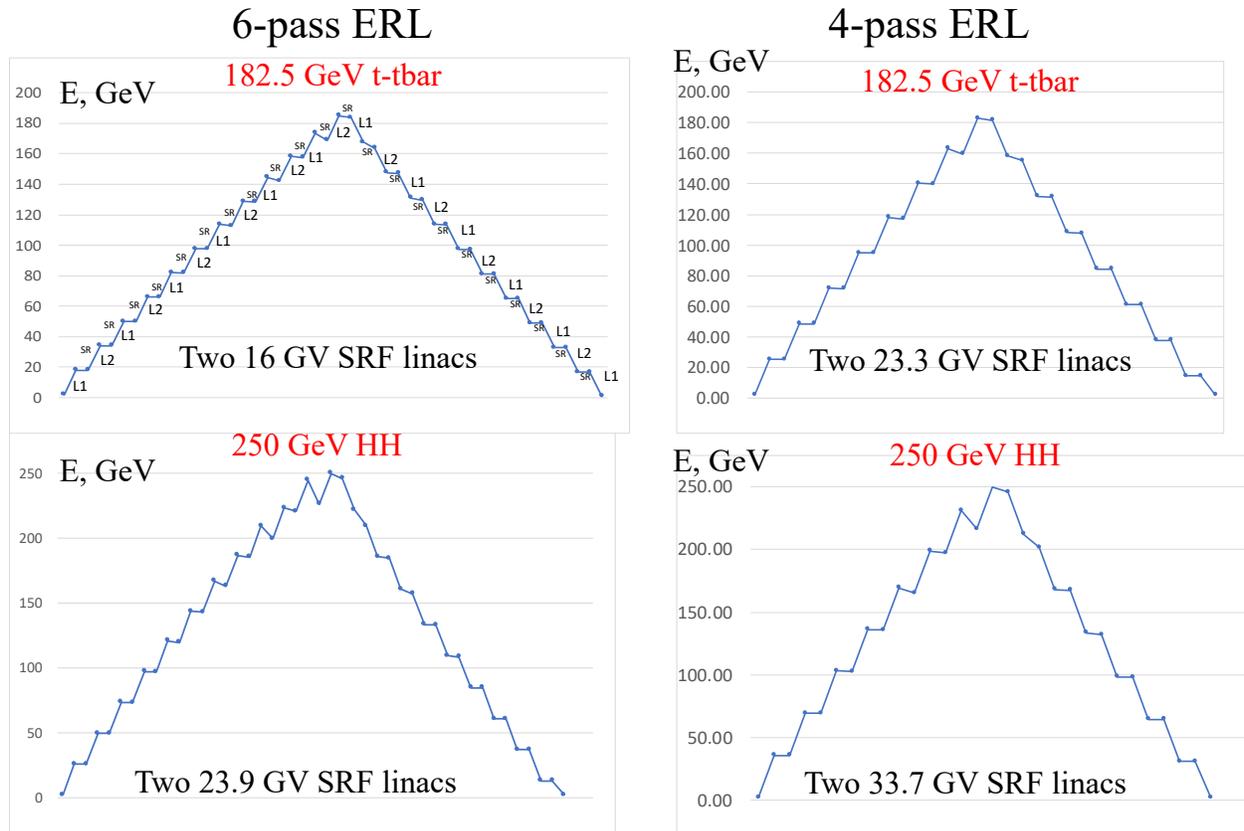

Fig. 4. Graphs of energy evolution in FCC ee 4-path and 6-path ERLs including SR losses. In the top-left graph the reason for the energy change in indicated: SR – loss from synchrotron radiation, L1 and L2 – energy gain/loss in linacs one and two.

Figure 5 shows a possible layout of the transport arcs in the FCC tunnel. It is an array of small gap electromagnets with round vacuum manifolds for pumping and absorbing the synchrotron radiation, which are naturally located at the outer side of the ring. To avoid parasitic collisions, electrons and positrons have separate magnetic systems. To maintain synchronism, they have alternate passes in the inner and the outer arcs. On the IR side, the top energy beams propagate through a dedicated line that passes through the IRs while the other beams by-pass the detectors. The beams from separate arcs are merged to pass through the linacs and then again separated by magnetic systems called combiners and separators (see for example [4-5]).

While looking rather elaborate, when build with small gap magnets, as shown in Fig. 6, such systems can be relatively inexpensive and have low power consumption. The most important feature of the arcs at top energy is the preservation of the beam emittance close to that provided by the cooling rings while keeping the synchrotron radiation losses as low as possible. This can be achieved by keeping the dipole field as low as possible and the filling factor as high as possible.

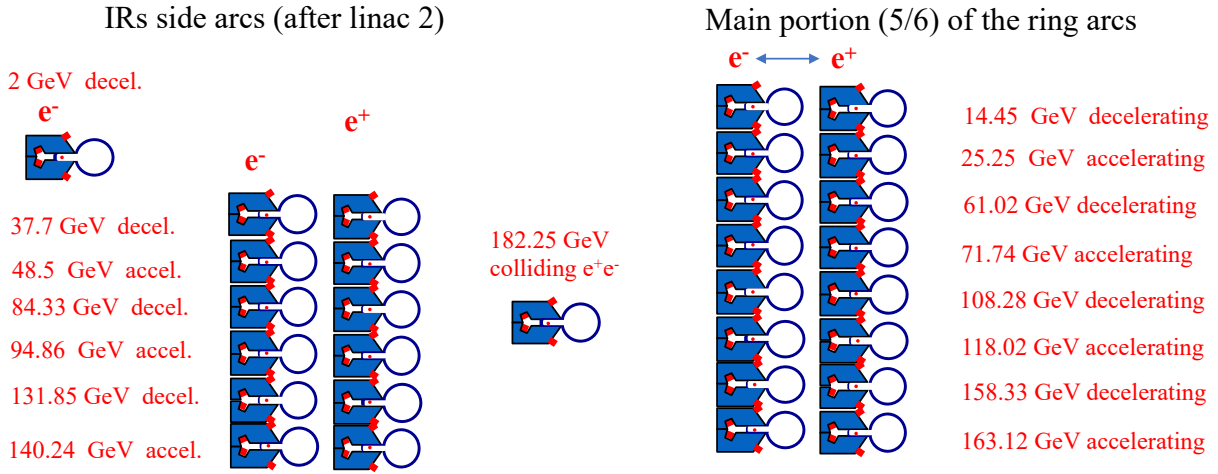

Fig. 5. Possible layout of the arcs for 4-pass ERLs with small gap electromagnets, similar to an early eRHIC linac-ring design [4]. The energies of the beams are shown for a top energy of 182.25 GeV (t-tbar).

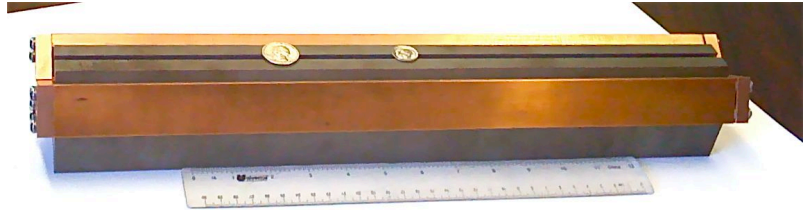

Fig. 6. Electro-magnets with a small 5 mm gap prototyped for the eRHIC linac-ring design and tested at 0.43 T [9-10]. FCC ee needs dipoles with a magnetic field of up to 0.04 T and could be driven by a coil with low current.

The most economical way to satisfy these contradicting requirements is to use combined function dipole magnets with a fixed magnetic but alternating gradient, e.g. a FODO lattice with added sextupole components to compensate for the natural chromaticity in the arcs. In this case the magnetic field can be kept constant at the beam's orbit, gaps between magnets can be small and the filling factor is large. The very small vertical emittance makes it natural to select small gap magnets. Since the dipole field is very low and the gaps are small such electro-magnets would consume very little electric power.

The preservation of the beam emittance is less trivial. The growth of the transverse emittance (diffusion) is proportional to the beam energy to the power seven, e.g. the highest energy passes have to have very low emittance optics. A straight forward estimation of emittance growth at the energy of 250 GeV shows that using the proposed FODO lattice with a 16-meters period (two 8-meter magnets) and a phase advance of 90 degrees provides sufficient emittance preservation to satisfy the requirements specified in Table 1. Conditions for the lower FCC ee energy operations are then also satisfied.

Another potential emittance degradation can come from errors caused by ejection errors (jitter) from the cooling rings. While this issue is also present in linear colliders there is no way for the correction signal to catch up with the beams moving at the speed of light. In contrast with a linear collider, for an ERL the transverse position errors caused by the pulsed ejector magnets can be corrected at the first arc when the beam travels around the FCC: the position and angle can be detected at the arc entrance and corrected at its exit, as is done in synchrotron light sources with similar small emittances. If necessary, this process can be repeated at every path around the ring.

Finally, it is important to note that the beams have to be extracted from linacs at an energy of 2 GeV at the end of deceleration and that the location of the extraction would depend on the operational energy of the FCC. The cooling rings are a critical part of the ERL-based FCC ee scenario. They are used to accumulate energy-recovered electrons and positrons and cool them down to the natural longitudinal and transverse emittances, which is typical for 4th generation light sources [7-8]. Table 3 gives the main parameters of the cooling rings for FCC ee. Achieving the listed beam emittances would require damping wigglers [7] as well as bunch stretching using flat RF buckets created by harmonic cavities.

Table 3. Typical cooling ring parameters required for ERL-based FCC ee

| Beam energy | 2 | GeV |
|---|---|---|
| Magnetic field, B | 1 | T |
| Energy loss rate | 1500 | GeV/sec |
| Filling factor | 0.67 | |
| Damping time | 0.002 | sec |
| Beam cooling time | 0.004 | sec |
| Ring circumference | 900 | m |
| Revolution frequency | 0.33 | MHz |
| Normalized emittance, hor | 4 | μm rad |
| Normalized emittance, vertical | 8 | nm rad |

Furthermore, both lepton beams should undergo compression during the first pass around the FCC tunnel as well as decompression during the last pass prior to ejection to the cooling ring. Long bunches, requiring subsequent compression, have a relatively low peak current in the cooling rings to mitigate IBS, while the de-compression will provide a reduction of the energy spread accumulated in the returning beam to fit into the energy acceptance of the cooling rings. Using the low-energy passes of the ERL for the compression and decompression will provide for a large value of the longitudinal dispersion $R_{56}$, while maintaining low emittance growth. This stretching and compression will require additional RF gymnastics, such as chirping beam energy and compensating the energy chirp after the bunch compression/decompression.

*Beam-beam effects.* For simplicity we used values of the $\beta_{x,y}^*$ similar to that in the current ring-ring design [6], but without using crossing angles. In other words, we assume head-on collisions of the bunches. We also assumed to use flat beams with a similar ratio of about 1000 between the vertical and horizontal emittances of lepton beams as in the ring-ring design, but we assumed operating with very low emittances typical for 4th generation light sources, e.g. ~ few nm rad horizontal geometrical emittance at 2 GeV. Table 4 compares beam parameters of ERL and ring-ring FCC ee options at t-tbar energy with ILC and CLIC parameters at similar energies[5].

---

[5] Parameters for ILC and CLIC are taken from presentation of Daniel Schulte [20]. We used beam energies closest to the top FCC ee ring-ring design for this comparison.

Table 4. Comparison of the ring-ring and ERL-based FCC ee with ILC and CLIC.

| Parameter | Ring-Ring | ERL-ERL | ILC @250 GeV | CLIC @ 190 GeV |
|---|---|---|---|---|
| Horizontal norm ε, μm rad | 518 | 8 | 10 | 1 |
| Vertical norm ε, nm rad | 964 | 8 | 35 | 30 |
| Horizontal β, m | 1.0 | 1.0 | 0.1 | 0.8 |
| Vertical β, mm | 2.0 | 2.0 | 0.5 | 0.1 |
| RMS bunch length, mm | 2.0 | 2.0 | 0.3 | 0.07 |
| Beam collision rate, kHz | 116.9 | 15.0 | 6.5 | 17.6 |
| Bunch charge, nC | 46.2 | 22.5 | 3.2 | 0.8 |
| Beam current, mA | 5.40 | 0.34 | 0.021 | 0.015 |
| Particle energy loss, GeV | 9.2 | 14.8 | 250.0 | 190.0 |
| Beam losses, MW (two beams) | 100.00 | 9.98 | 10.40 | 5.55 |
| Energy spread in IP, % | 0.18 | 0.16 | - | - |
| Dx/Dy | N/A | 0.2/143 | 0.3 / 24.3 | 0.24 / 12.5 |
| Crossing angle | YES | NO | YES | YES |

It is well-known from linear collider studies that at high beam energies the most dangerous effect causing energy spread is beamstrahlung, e.g. synchrotron radiation in strong EM field of the opposing beam during collision. It is also well known – and used in the ring-ring FCC ee design – that the best mitigation for beamstrahlung is the use of flat beams with a large aspect ratio between the horizontal and vertical beam sizes. For flat beams with $\sigma_x \gg \sigma_y$ one can calculate the RMS energy spread induced by the beamstrahlung to be

$$\sigma_\gamma = \frac{4}{9}\sqrt{\frac{\pi}{3}} N^2 \frac{r_e^3}{\sigma_x^2 \sigma_z} \gamma^2 \qquad (6)$$

We use this formula to estimate the RMS beam energy spread at the IRs (Table 4) and found that it is comparable to the energy spread for the ring-ring design.

The other important effect in our proposed design is growth of the beam emittances resulting from a single beam-beam collision. In contrast with ring colliders, but the same as for linear colliders, beam collisions in ERL-based colliders are described by the disruption parameter:

$$D_{x,y} = \frac{N_e}{\gamma_e} \frac{2r_e}{(\sigma_x + \sigma_y)\sigma_{x,y}} \sigma_z; \quad \sigma_{x,y} = \sqrt{\beta_{x,y} \varepsilon_{x,y}}, \qquad (7)$$

which represents the strength of the focusing by the opposite beam during collisions. It is expected that large disruption parameters would result both in pinching of beam sized as well as in transverse emittance growth. We conducted preliminary studies of these effects in strong-strong collision of two 250 GeV beam with parameters shown in Tables 1 and 2. We used similar technique used for our eRHIC beam-beam studies by adjusting location of the "geometrical" beam waists – so call s* - to minimize the pinch effect and the emittance blow-up. Nonlinear time-dependent EM field induced by cooling beam introduce very interesting dynamics of the particles in the phases space. Horizontal motion is partially frozen during the collision, which occurs at the mm-scale of the bunch length, while horizontal motion is if defined by $β_x$ of 1 meter. In contrast, high vertical disruption parameter ($D_y=121$ in this case) results in up to two vertical oscillations the during collisions. Furthermore, flat colliding beam make vertical oscillations function of horizontal position of the particle: it is strong in the beam center and fades away at the beam's edges. Fig. 7 illustrated this dependence. Beam was split in 201 slices along the bunch length from -3 to +3 RMS bunch length for illustrating dependence on particle positions in the colliding bunches.

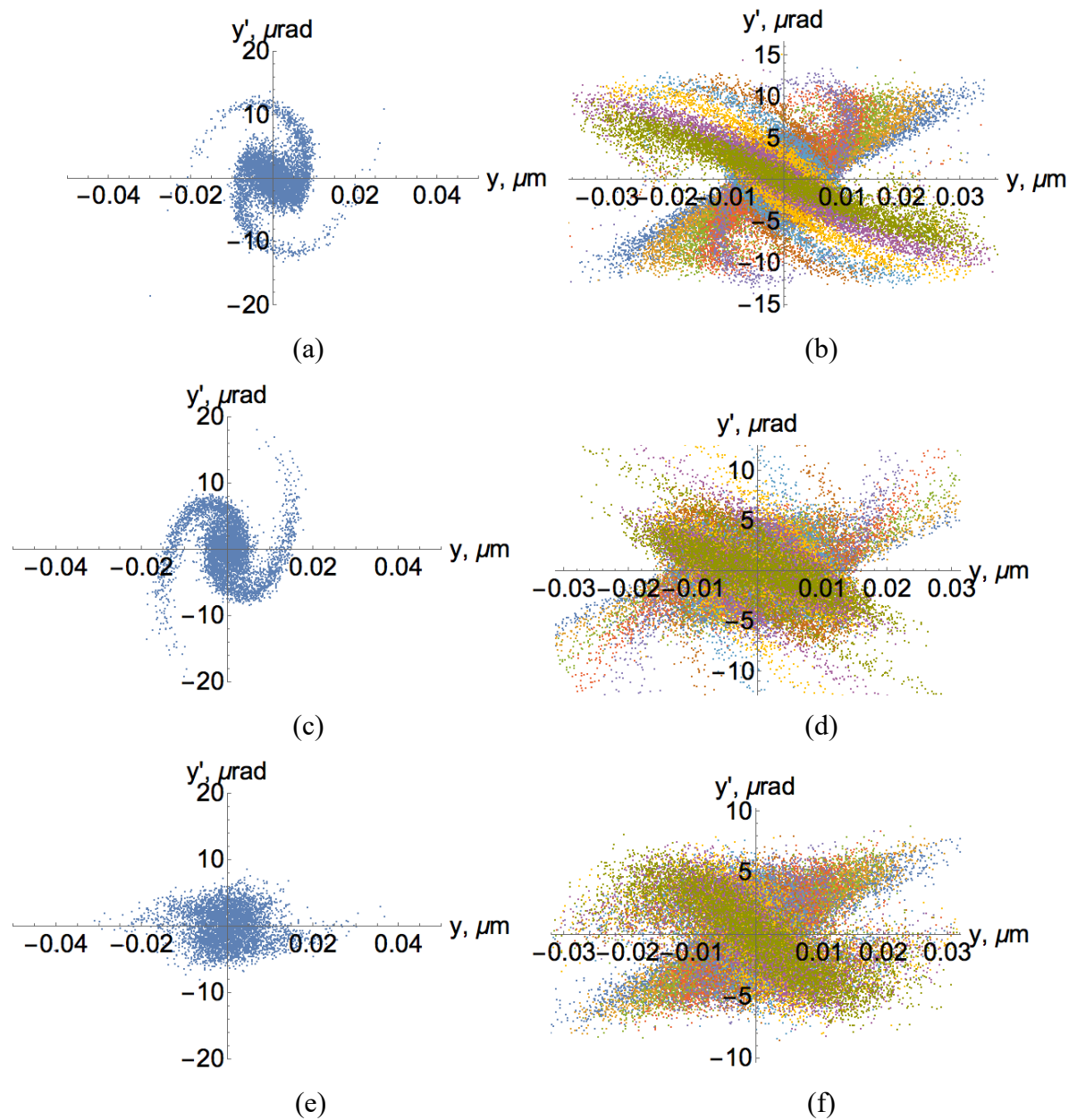

Fig. 7. Beam distribution in vertical phase space after the collision. Distribution of the central slice is on the left and combinations of 10 covering evenly $-3\sigma_z < z < 3\sigma_z$, is on the right: (a-b) are for center parties at x=0; (c-d) is for x= $\sigma_x$, (e-f) is for x= $2\sigma_x$. Horizontal axis is vertical coordinate in micrometers and vertical axis is vertical angle of the particle in microradians.

Fig. 7 clearly shows that vertical motion is strongly nonlinear. Still, our simulations using geometrical settings of *β\*=1.7 mm* as *s\*=-2.8 mm* show that during collision with vertical disruption parameter $D_y=121$ vertical emittance grows less than a factor two for regular focusing. This value reduces to 1.86 when we assumed crab-focusing.

For particle close to the beam axis $D_y=121$ corresponds to approximately two vertical betatron oscillations during the collision. In contrast, particles far from the axis (see Fig. 7 (e,f)) a weakly affected by the collision. Our results also show that with proper choice of the optics emittance growth is a very weak function of the disruption parameter – in other words the main "damage" to beam emittance happens during first oscillation occurring with disruption parameter ~ 25. Further increase of the disruption parameter up to ~ 200 causes more rotation of "spiral galactic of particles" with tails at large amplitudes wrapping around the beam core.

These results also indicate a possibility of colliding beams in multiple IRs to increase total luminosity of the collider – but this option should be carefully simulated to account for possible drawbacks. It is important to note that it is possible to increase blow-up of vertical beam emittance significantly by using strongly mismatched optics, e.g. with significant deviations from parameters used in our simulations.

We also simulations evolution of the beam sizes and instantaneous luminosity during the collision: see Figs. 8 and 9. Fig. 8 show evolution of the beam sizes of various z-slices for particles located close ti x=0: one can easily see so-called "pinch effect originating from the focusing imposed by the opposite beam compressing beams to vertical RMS size ~ 2 nm from initial value of ~ 5 nm.

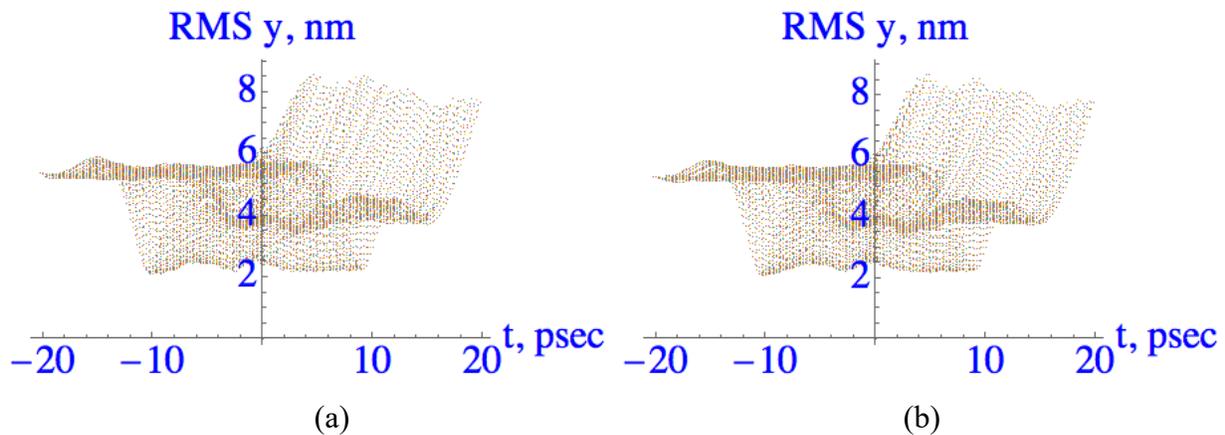

Fig. 8. Evolution of RMS beam sizes for various slices during the collision: (a) for electron beam and (b) for positron beam. Envelopes are nearly identical for two beams.

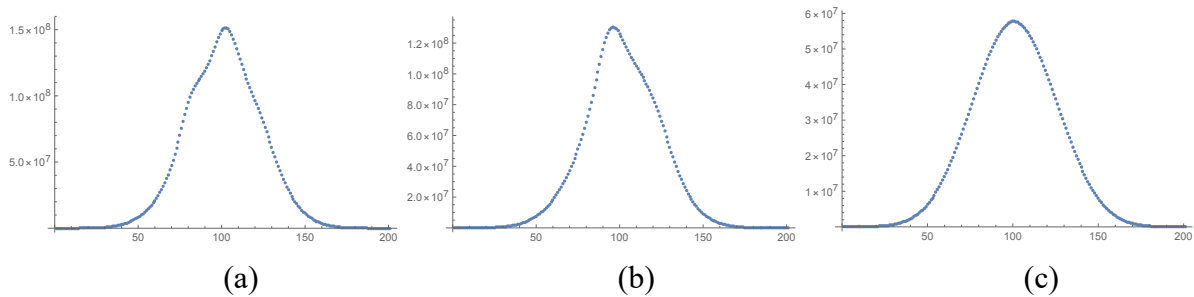

Fig. 9. This graph illustrates instantaneous luminosity (in arbitrary units) for the portion of the beams located at (a) $x=0$; (b) $x=\sigma_x$ and (c) ) $x=2\sigma_x$. Horizontal axis is in time steps of 20 fsec used in this simulation.

The pinch effect leads to reduction of the beam size and corresponding increase in the luminosity. Fig. 9 shows instantaneous luminosities during beam's collision. Unusual (asymmetric) dynamics of the instantaneous luminosity (for the core of the beam) is related the pinch-effect as well as to the emittance growth.

Our simulations were aiming on demonstrating possibility of modest (~ 2 fold or less) increase of the vertical emittance during collisions. Hence, we did not attempt to optimize the collisions and further improvements, not only in the beam emittances but also in the luminosity, are possible.

*Conclusions and acknowledgements*. In this short paper we discuss a possibility of ERL based future circular electron positron collider (FCC ee) and compare its performance with previously explored in depth ring-ring design [1-2] and linear colliders ILC and CLIC. We found that at c.m. energies from160 GeV to 500 GeV the ERL-based FCC ee promises higher energy reach and higher luminosity while consuming significantly less power than the current ring-ring design.

While ERL-based FCC ee has similar IR parameters and flat-beam geometry used in the ring-ring design, it allows for use of much smaller beam emittances. While such emittance can be generated in $4^{th}$ generation light sources, they cannot be used in the ring-ring design because the violation of the beam-beam limit will result in instantaneous emittance blow-up or even in loss of the beams. In contrast, in the ERL-based FCC ee beams are energy recovered after each collision and cooled down to naturally low emittances in the 2 GeV cooling rings. We demonstrated in our simulation that in a single IR collision beam's vertical emittance will increase by less than a factor of two and, therefore, can be comfortably energy recovered and recycled in the cooling ring. As we mentioned previously, both electron and position beams can be polarized.

Tiny losses in the recycled beams – which can come from the scattering on the residual gas or IBS – will be replenished by 2 GeV injectors of electron and position beams. The latter is the main difference between linear colliders and ERL-based FCC ee – in ERL-based collider we not only recover significant portion of the beam energy but also recycle the beams. In current designs of linear colliders, the beam dumped at the top energy and sources of fresh electrons and positrons with full beam current are needed to support them.

Up to date we did not find any showstoppers for ERL version of FCC ee. In contrast with linear colliders, the transverse position jumps and jitters caused by pulsed ejector magnets or vibrations can be corrected as the beam passes around the FCC: the position and angle can be detected at the arc entrance and corrected at the exit. Geometric emittances and transverse beam sizes are very small making it natural to use low-cost small gap magnets. Such combined-function magnets with alternating gradients (bend-quadrupole channel) have extremely high energy acceptance measured in units of energy [19] and provide for a constant bending magnetic field with ~ 90% packing factor. The later gives about a 35% savings in the synchrotron radiation power.

In short, the ERL option, in combination with 2 GeV cooler rings, could be advantageous for FCC ee high energy operation. It promises significant, 6 to 10 times, reduction in the required RF power while delivering higher luminosities at top energies. There should be no problems with beam stability in ERL: a very low average current and the use of modern higher-order mode (HOM) dampers [16-18] will be sufficient to keep the beams stable in such an ERL.

It is important to notice that the ERL scheme does not have advantages over the ring-ring design at the lowest FCC ee energy of 46.5 GeV.

Clearly, detailed in-depth studies – comparable with those done for the ring-ring design - are needed to fully validate this ERL-based concept. Specifically, full range start-to-end simulations are needed for a full validation of a such design. Nevertheless, if valid, the ERL-based FCC ee would have additional advantages when compared with alternative designs:

1. Reducing the SR power will extends the FCC ee life-cycle. As it was demonstrated in LEP, synchrotron radiation with MeV photons degrades the surrounding hardware
2. Polarized beams can be used in an ERL-based FCC ee
3. ERLs can serve as the lepton part of a future FCC eh collider.

Authors would like to thank Dr. Frank Zimmerman and FCC-design team at CERN for opportunity to present and discuss this option at their working meeting. Vladimir Litvinenko would like to acknowledge support by NSF grant PHY-1415252 "Center for Science and Education at Stony Brook University".